\documentclass[twocolumn,preprintnumbers,showpacs,aps,
               amsmath,amsfonts]{revtex4}
\usepackage{epic,eepic}
\usepackage{hyperref}

\newcommand{\lp}{\left}
\newcommand{\rp}{\right}
\newcommand{\be}{\begin{eqnarray}}
\newcommand{\ee}{\end{eqnarray}}
\newcommand{\beq}{\begin{equation}}
\newcommand{\eeq}{\end{equation}}
\newcommand{\ba}{\begin{array}}
\newcommand{\ea}{\end{array}}

\DeclareMathOperator{\sgn}{sgn}

\begin{document}


\preprint{CALT-68-2464}

\title{Serial composition of quantum coin-flipping, \\
and bounds on cheat detection for bit-commitment}

\author{Carlos Mochon}
\email{carlosm@theory.caltech.edu}
\affiliation{Institute for Quantum Information, 
California Institute of Technology,
Pasadena, CA 91125, USA}

\date{October 1, 2004}

\begin{abstract}
Quantum protocols for coin-flipping can be composed in series in such a
way that a cheating party gains no extra advantage from using entanglement 
between different rounds.
This composition principle applies to coin-flipping protocols with cheat 
sensitivity as well, and is used to derive two results:
There are no quantum strong coin-flipping protocols with cheat sensitivity
that is linear in the bias (or bit-commitment protocols with linear cheat 
detection) because these can be composed to produce strong coin-flipping 
with arbitrarily small bias.
On the other hand, it appears that quadratic cheat detection cannot be composed
in series to obtain even weak coin-flipping with arbitrarily small bias.
\end{abstract}



\maketitle

\section{Introduction}

Coin-flipping is the cryptographic problem where two mutually distrustful 
parties try to agree on a random bit. This is to be accomplished by sending 
sequential messages over a communication channel.

The goal is that, if both players are
honest and follow the prescribed protocol, they both output the 
same bit, which is uniformly random. The restriction is that, if 
either party is dishonest, then the bit output by the honest player 
must still be approximately random. The figure of merit for a coin-flipping
protocol is the bias, defined as the maximum increase in the probabilities of
the honest player's outputs that a cheating party can achieve.

There is a weaker version of the above problem, where one party wants to
obtain outcome zero, and the other party outcome one. 
This corresponds more closely
to the colloquial idea of coin-flipping. In this case, the protocol need only 
prevent a cheater from biasing the coin in favor of
his desired outcome. The two versions
of the problem are known respectively as strong and weak coin-flipping.

In the classical setting the problem is impossible without some additional 
restrictions on the computational power available to each party. Nonetheless,
in the quantum setting, where the parties can process quantum information and
communicate over a quantum channel, the goal can be partially achieved.

Ambainis \cite{Ambainis2002} and Spekkens and Rudolph \cite{Spekkens2001} have 
constructed strong coin-flipping protocols with a bias $\epsilon=\frac{1}{4}$.
However, it was proven by Kitaev \cite{Kitaev} (and summarized in 
Ref.~\cite{Ambainis2003}), that all quantum protocols for strong coin-flipping
must have a bias of at least 
$\epsilon=\frac{1}{\sqrt{2}}-\frac{1}{2}\simeq 0.21$.

Quantum weak coin-flipping is a little more promising. The best known protocol
is by Spekkens and Rudolph \cite{Spekkens2002} and achieves a bias of
$\epsilon=\frac{1}{\sqrt{2}}-\frac{1}{2}$. The only known bound is due to
Ambainis \cite{Ambainis2002}, and states that the number of rounds
must grow at least as $\Omega(\log \log \frac{1}{\epsilon})$. 
In particular, to obtain
arbitrarily small bias, we must have protocols with an ever increasing 
number of rounds.

Unfortunately, analyzing quantum protocols with a large number of 
rounds is often difficult. One approach to obtaining a weak coin-flipping 
protocol with arbitrarily small bias could be to take a quantum coin-flipping 
protocol with a fixed number of rounds, and compose it in series with itself
to obtain a better coin-flipping protocol. It is well known that quantum
coin protocols compose well in series, and an argument for this is given
in Sec.~\ref{sec:sercom}.

Unfortunately, serially composing standard coin-flipping protocols does not
decrease the overall bias \cite{Santha86}. 
However, quantum mechanics is good at detecting
state disturbance, and other deviations from a protocol. It is therefore 
possible to construct coin-flipping protocols with cheat sensitivity, where
a dishonest player may be able to cheat by a significant amount, but only
at the risk of getting caught by the honest player. 
Cheat sensitive protocols can produce improved coin-flipping protocols
when composed in series under certain conditions.

Quantum protocols with cheat sensitivity have been constructed by
Spekkens and Rudolph 
\cite{Spekkens2002} where players cheating by large amounts will get 
caught, but at least one party can bias the coin by a small amount without 
being detected. Aharonov \textit{et al}.
\cite{Aharonov00} have a protocol with quadratic cheat detection 
on one side, but no cheat detection on the other side. Hardy and Kent 
\cite{Hardy99} devised a protocol where neither player may cheat by any amount
without risking detection, but the functional form of the cheat detection
was not determined, though it is known to be quadratic or worse.
Functional forms of cheat detection were also discussed in 
Ref.~\cite{Spekkens2003}.

In the present paper, we will analyze the serial composition of cheat
sensitive coin-flipping protocols. We shall treat the cheat sensitive
protocols as oracles or black boxes, with a cheat sensitivity that is
given by a function of the target bias. This will lead to our two main
results:

In Sec.~\ref{sec:quad} we show that quadratic cheat detection protocols,
where the probability of getting caught is proportional to the square of the
bias, are a fixed point of serial composition, at least to leading order.
This means that no matter how many times the protocol is composed with itself,
the amount of cheat detection remains approximately the same. 
Because most known
cheat sensitive protocols are quadratic or worse, this result is evidence 
that serial composition may not be useful to obtain weak coin-flipping.

In Sec.~\ref{sec:lin} we show that linear cheat detection cannot exist for 
strong coin-flipping. This is done by composing the linear cheat detection
to obtain a strong coin-flipping protocol with arbitrarily small bias,
in violation of Kitaev's lower bound. Note that the second result only
uses serial composition as a tool for the proof, and the result holds 
for all cheat sensitive strong coin-flipping protocols.

The result of Sec.~\ref{sec:lin} also applies to bit-commitment, which is a
cryptographic protocol related to coin-flipping. Coin-flipping can be 
constructed out of bit-commitment as follows: first Alice commits a bit
to Bob, then Bob announces a random bit, then Alice reveals her bit and the
coin outcome is the \textsc{xor} of the two bits. Cheat detection as a function
of $\epsilon$ can be defined in a way similar to Ref.~\cite{Aharonov00}.
For Alice, $\epsilon$ is the amount by which she can change the probabilities
associated with the committed bit, whereas for Bob, 
$\epsilon$ is the additional
probability of guessing Alice's committed bit correctly. Because
linear cheat detection in bit-commitment can be used to produce a strong
coin-flipping protocol with linear cheat detection, it is also ruled out
as a possible quantum protocol.

\section{\label{sec:sercom}Serial Composition of Quantum Coin Protocols}

Protocols for coin-flipping can be naturally composed in series to 
obtain new coin-flipping protocols.
What is surprising at first is that quantum protocols for coin-flipping 
compose serially in such a way that a cheating party does not get any
unexpected advantage by using entanglement. We shall prove this below, 
but let us first define carefully what we mean by unexpected advantage.

Let $\mathcal{P}$ be any quantum coin-flipping protocol. At the end of the
protocol each player will output one of $\{0,1,C\}$ where the last entry
denotes the output when one player catches the other player cheating. Let
us assume that Alice is honest and Bob is cheating. For each cheating
strategy employed by Bob, there will be a triple of probabilities 
$(P_0,P_1,P_C)$, one for each of Alice's possible outputs. Let 
$\Omega_A(\mathcal{P})$ be the set of all such attainable triples. 
Clearly $(1/2, 1/2, 0)\in\Omega_A(\mathcal{P})$ since Bob can always play 
honestly. If the protocol does not allow Bob to fully bias toward $1$ then
$(0,1,0)\notin\Omega_A(\mathcal{P})$. Some protocols may not have any cheat 
detection, in which case $P_C$ will always be zero.
For honest Bob and cheating Alice there is a similarly defined 
$\Omega_B(\mathcal{P})$.

Assume we take the protocol $\mathcal{P}$ and run it many times in series.
We are interested in proving that a cheating player, say Bob, does not
obtain any extra cheating power by using entanglement between different
rounds. That is, that for every round $j$, independently of previous outcomes
and strategies used by Bob, any strategy that Bob employs will make Alice 
output based on a triple of probabilities in $\Omega_A(\mathcal{P})$.

Clearly Bob can vary his strategy in each coin-flip round, and even
base his strategy on the outcomes of the previous coin-flips. However,
when flipping $N$ coins in series, Bob cannot obtain an outcome of all
ones with a probability greater than $(P_{1,\text{max}})^N$, where
$P_{1,\text{max}}$ is the maximum value of $P_1$ over
any triple in $\Omega_A(\mathcal{P})$.

The reason that quantum coin-flipping can be serially composed stems from the 
following conditions which are always imposed on coin-flipping protocols:
\begin{enumerate}
\item There is always at least one honest party.
\item The details of the protocol, which can be described in terms
of fixed unitaries acting on a fixed initial state, are known to all parties.
\item The protocol begins in an unentangled state, a condition which
can be enforced by the honest party.
\end{enumerate}
\noindent
The first condition arises because no constraint is imposed on the case
when both parties are cheating, and therefore the case never needs 
consideration. The third condition is always imposed to avoid trivial 
protocols, since establishing correlations is the goal of a
coin-flipping procedure.

The proof of composition is fairly simple. 
At the beginning of the $k^{th}$ round, the 
cheater will be unentangled from the honest player. The honest player
has erased her Hilbert space and reset it to the initial state of the 
protocol. All that the cheating player has left over from the previous 
rounds is some quantum state in his Hilbert space. However, because the honest
protocol is public, the cheating player knows exactly the state of his
Hilbert space (which may be a mixed state, caused by the honest player erasing
her Hilbert space). 

Now assume that he could, with the help of this state, force the honest player
(say Alice) to output with probabilities not in $\Omega_A(\mathcal{P})$. 
Then with the same state, he could obtain the same results in the first round
or even when protocol $\mathcal{P}$ is used in a one-shot run, contradicting
the definition of $\Omega_A(\mathcal{P})$. To put it another way, Bob can 
simulate in his private Hilbert space the first $k-1$ rounds and start playing
the first round from that point, but this clearly can give him no extra 
advantage.

The conclusion of this section is that, given a quantum protocol for
coin-flipping, we may treat the protocol as a black box when composing
it in series with itself (or even with other coin-flipping protocols)
without worrying about entanglement between rounds. We shall use this fact
to derive two interesting results.

\section{\label{sec:quad}Quadratic cheat detection is a fixed 
point of coin-flipping}

Because quantum coin-flipping composes in series, it is tempting to try to
use a classical game layer on top of a known quantum coin-flipping protocol
in order to reduce its maximum bias. That is, we wish to construct a two
player classical game that uses the quantum coin as a black box. 
The game could be, for instance, flipping a coin $N$ times, 
and the party that wins a majority of coin tosses wins the game.

The ideal goal for this process would be to produce a weak coin-flipping 
protocol with arbitrarily small bias. While it is known that games built
out of standard coin-flipping protocols can never reduce the maximum bias,
the situation is different when cheat detection is available. Especially
in the case of weak coin-flipping, where an honest player may declare himself
the winner if he detects the other party cheating, there are certain
black-box protocols that can be used to produce arbitrarily small bias.
The question is how much cheat detection is needed in order for successive
compositions to improve a coin-flipping protocol?

We will be interested in protocols with symmetric, monomial cheat detection.
Let $\mathcal{P}$ be a protocol where both parties have equal cheating 
opportunities 
(i.e.,~$\Omega_A(\mathcal{P})=\Omega_B(\mathcal{P})\equiv\Omega(\mathcal{P})$)
and such that all probabilities $(P_0,P_1,P_C)\in\Omega(\mathcal{P})$ 
have the form:
\be
P_0 &=& (1- P_C)\lp(\frac{1}{2}+\epsilon\rp),\\
P_1 &=& (1- P_C)\lp(\frac{1}{2}-\epsilon\rp),\\
P_C &\geq& a |\epsilon|^b,
\label{eq:Pc}
\ee

\noindent
which can be viewed as a function of a parameter $\epsilon$ that is 
controlled by the cheating party. The constants $a>0$ and $b\geq 0$ 
denote the amount of cheat detection. Strictly speaking, $P_C$ should also be 
considered a second parameter that can be controlled by the cheater, as he
can always use a less optimal cheating strategy. In practice, a cheater
will always minimize $P_C$ for a given bias, and therefore we may assume
equality in Eq.~(\ref{eq:Pc}). We shall call the case $b=1$
linear cheat detection, and the case $b=2$ quadratic cheat detection. 

When building games out of cheat sensitive coin-flips, the outcome
of the entire game will be ``cheating'' if in any individual coin-flip
a cheating outcome was obtained. To leading order in epsilon, 
the composite game will have a cheat sensitivity of the same form,
with the same coefficient $b$, but in general a different $a$.
 When $b$ is small, successive compositions increase 
$a$ thereby producing a more cheat sensitive protocol, whereas in the large 
$b$ regime composition has the opposite effect. We intent to prove in this 
section that $b=2$ is a fixed point of coin-flipping. That is,
for all possible games that use a coin with quadratic cheat detection
as a black box, the resulting protocol has exactly the same amount
of cheat detection to leading order in epsilon.

\begin{figure}[tb]
\setlength{\unitlength}{0.00083333in}
{
\begin{picture}(2863,1475)(-600,-50)
\path(12,675)(912,1125)
\path(12,675)(912,225)
\path(912,225)(1212,75)
\path(912,225)(1212,375)
\path(1212,375)(1512,225)
\path(1212,375)(1512,525)
\path(1212,975)(1512,1125)
\path(1212,975)(1512,825)
\path(912,1125)(1212,975)
\path(912,1125)(1212,1275)
\put(12,675){\circle*{50}}
\put(912,225){\circle*{50}}
\put(912,1125){\circle*{50}}
\put(1212,375){\circle*{50}}
\put(1212,975){\circle*{50}}
\put(1587,1125){\makebox(0,0)[l]{0}}
\put(1587,825){\makebox(0,0)[l]{1}}
\put(1587,525){\makebox(0,0)[l]{0}}
\put(1587,225){\makebox(0,0)[l]{1}}
\put(1287,75){\makebox(0,0)[l]{1}}
\put(1287,1275){\makebox(0,0)[l]{0}}
\put(-80,750){\makebox(0,0)[r]{$^{P_W=\frac{1}{2}}$}}
\put(-80,575){\makebox(0,0)[r]{$^{\Delta=\frac{1}{2}}$}}
\put(820,200){\makebox(0,0)[r]{$^{P_W=\frac{1}{4}}$}}
\put(820,25){\makebox(0,0)[r]{$^{\Delta=\frac{1}{2}}$}}
\put(820,1325){\makebox(0,0)[r]{$^{P_W=\frac{3}{4}}$}}
\put(820,1150){\makebox(0,0)[r]{$^{\Delta=\frac{1}{2}}$}}
\put(1200,525){\makebox(0,0)[r]{$^{P_W=\frac{1}{2}}$}}
\put(1175,375){\makebox(0,0)[r]{$^{\Delta=1}$}}
\put(1200,925){\makebox(0,0)[r]{$^{P_W=\frac{1}{2}}$}}
\put(1175,775){\makebox(0,0)[r]{$^{\Delta=1}$}}
\end{picture}
}
\caption{``Best two out of three'' game tree.
Binary nodes are labeled by the probability of winning for an honest 
player trying to obtain outcome $0$.}
\label{fig:2oo3}
\end{figure}
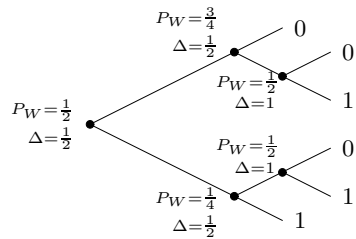

We begin by describing the set of all possible games that employ a cheat 
sensitive coin-flip as a black box. These can be put in correspondence
with the set of binary trees, where the leaf nodes are labeled by either
zero or one. For instance, the tree corresponding to the 
``Best two out of three'' game is depicted in Fig.~\ref{fig:2oo3}.
Each binary node corresponds to a coin flip, with up corresponding to the
outcome zero and down to the outcome one. The leaves are endpoints of the
game and their labels correspond to the final game outcome at that node.

Other examples of games that produce coin-flips include ``first outcome 
that is repeated $N$ times'' and ``first outcome to occur a total of 
$N$ times more than the other'',
both of which correspond to trees of infinite length. Allowing for trees
of infinite length also accounts for all games that have a tie outcome,
after which the game is restarted. For trees of infinite length we impose
the additional constraint that when playing honestly, the probability of
never reaching a leaf node is zero.

Let $x$ be a variable that ranges over the binary nodes of the 
tree. A general cheating strategy
is a function $\epsilon(x)$, which assigns a bias to each node where a 
coin-flip takes place. We also define $p_\epsilon(x)$ as the probability
of arriving at node $x$ given a cheating strategy $\epsilon(x)$. In
defining this quantity, we assume that if the cheater is caught at a
given node, the game stops, thereby reducing the probability of arriving
at the child nodes. In terms of these quantities we can write the total
probability of getting caught:
\be
P_{C,tot} = \sum_x a p_\epsilon(x) \lp|\epsilon(x)\rp|^b .
\ee

\noindent
To leading order in $\epsilon(x)$, we can replace $p_\epsilon(x)$ with the
probability of arriving at the node in the honest protocol. This can be 
written as $2^{-D(x)}$ where $D(x)$ is the depth of the node $x$, with the root
node having depth zero. The formula becomes:
\be
P_{C,tot} = \sum_x a 2^{-D(x)} \lp|\epsilon(x)\rp|^b .
\ee

To expand the total bias to leading order in epsilon, we only need to keep
track of terms that are linear in $\epsilon(x)$, and can therefore disregard
the multiplicative factor $1-P_C$ in the probabilities for obtaining zero or
one. Of course, we are assuming $b>1$ at this point, and will soon concentrate
on $b=2$.
The probability of the cheater obtaining his desired outcome, and winning
the game, satisfies a simple recursion relation:
\be
P_W(x) &=& \frac{1}{2} \lp( P_W(x^\uparrow) + P_W(x^\downarrow) \rp)
\nonumber\\
&& + \epsilon(x) \lp( P_W(x^\uparrow) - P_W(x^\downarrow) \rp),
\ee

\noindent
where $P_W(x)$ is the probability of winning having arrived at node $x$,
and $x^\uparrow$, $x^\downarrow$ are the two children of node $x$. At the
root node the probability of winning is
\be
P_{W,tot} = \frac{1}{2} + \sum_x 2^{-D(x)} \Delta(x) \epsilon(x),
\label{eq:constr}
\ee

\noindent
where $\Delta(x) = P_W(x^\uparrow) - P_W(x^\downarrow)$, which can be computed
at this point from the honest probability of winning.

The total bias for the game is given to leading order 
by $|\epsilon_{tot}| = P_W - 1/2$, where we have excluded cases where
the cheating benefits the honest player.
For a given total bias, the cheater will choose $\epsilon(x)$ to minimize
the probability of getting caught. The calculation to minimize $P_{C,tot}$
under the constraint of fixed $\epsilon_{tot}$ can easily be done using a
Lagrange multiplier, to obtain the result:
\be
\epsilon(x) = \sgn(\lambda \Delta(x)) 
\lp|\frac{\lambda \Delta(x)}{a b}\rp|^{\frac{1}{b-1}},
\label{eq:eps}
\ee

\noindent
where $\lambda$ is the Lagrange multiplier. We have allowed $\epsilon(x)$ 
to range over the reals, but will show below that for quadratic cheat 
detection the optimal solution satisfies the requirement 
$|\epsilon(x)|\leq 1/2$. 

To eliminate the Lagrange multiplier, we can substitute the expression for
$\epsilon_{tot}$ into the expression for $P_{C,tot}$ to obtain:
\be
P_{C,tot} = a_{new} \lp|\epsilon_{tot}\rp|^b,
\ee
\noindent
where
\be
a_{new} = a \lp( \sum_x 2^{-D(x)} \lp|\Delta(x)\rp|^{\frac{b}{b-1}} \rp)^{1-b}.
\ee

For the case $b=2$, the following lemma shows that $a_{new}=a$ for all
game trees:

\noindent
\textbf{Lemma:} For all game trees as described above, we have the equality
\be
\sum_x 2^{-D(x)} \Delta(x)^2 = 1.
\ee
\noindent
The lemma can be proven using a simple combinatorial argument that is 
presented for the interested reader in the appendix. 

From Eq.~(\ref{eq:eps}) we see that for $b=2$, the amount of cheating at node
$x$ is proportional to $\Delta(x)$. The constant of proportionality, as well as
the Lagrange multiplier, are fixed by Eq.~(\ref{eq:constr}). After eliminating
a factor of $1$ using the above lemma we obtain that the optimal strategy
is $\epsilon(x) = |\epsilon_{tot}| \Delta(x)$. This satisfies the intuition 
that the cheater will choose a larger bias on coin-flips that are more 
consequential. Furthermore, because $|\Delta(x)|\leq 1$, we have shown
that the optimal strategy is achievable (i.e., $|\epsilon(x)|\leq 1/2$ for
every $x$) whenever $|\epsilon_{tot}|\leq 1/2$.

For other values of $b$ near two, we can consider the derivative of 
$a_{new}$ with respect to $b$. For any fixed graph, this derivative
is negative. The conclusion is that for every graph, serial composition
of $b<2$ cheat detection leads to improvement to lowest order, whereas
in the $b>2$ case the cheat sensitivity worsens.

The above results complete our argument that quadratic cheat detection
is a fixed point for coin-flipping. The argument is only valid in the
regime where all biases, including the total bias, are small. Unfortunately,
because a protocol producing weak coin-flipping with arbitrarily small bias
would have to employ arbitrarily large trees, it is not sufficient
to simply take the small bias limit on a fixed tree.

In essence, associated with each weak coin-flipping protocol there is a 
function $P_C(\epsilon)$ indicating the minimum probability with which
a party will get caught cheating if they try to bias the coin by $\epsilon$.
Composing this protocol in series one obtains a new protocol
with function $P_{C,tot}(\epsilon_{tot})$. That is, serial composition
with a given game tree induces a map from the set of functions 
$P_C(\epsilon)$ to itself. We have shown that, independent of the tree,
if the original function behaves as $a\epsilon^2$ for small $\epsilon$, then
so will its image.

The fact that the coefficient of the quadratic term remains fixed 
under the map induced by every game tree is a peculiar and interesting feature.
It is indicative that serial composition of quadratic cheat detection may not
be useful for producing weak coin-flipping with arbitrarily small bias.
However, further research in this direction is needed in order to conclusively
settle the issue.

\section{\label{sec:lin}No-go theorem for linear cheat detection}

In this section we shall switch gears and focus on serial composition of 
linear cheat detection protocols. We shall prove that linear cheat detection 
can be serially composed to produce not only weak coin-flipping but also 
strong coin-flipping with arbitrarily small bias. Because of Kitaev's bound 
(discussed in Ref.~\cite{Ambainis2003}), the result in this section
proves that a strong coin-flipping protocol with linear cheat detection
cannot exist.

The result of this section only applies to strong coin-flipping
schemes with cheat sensitivity as described in Eq.~(\ref{eq:Pc}).
An alternative not covered by the proof in this section is the case
where Alice can force outcome 1 (and Bob can force outcome 0) without getting 
caught, which corresponds to cheat sensitive weak coin-flipping.
Because weak cheat sensitivity
can be simulated by strong cheat sensitivity, the result of the previous
section applies to weak cheat sensitivity as well. However, the opposite is
not true, and the results of the present section do not apply
to linear cheat sensitivity in weak coin-flipping.

Since strong coin-flipping can be constructed
out of bit-commitment, the result of this section can also be applied to 
cheat sensitive bit-commitment protocols. In fact, the only known 
coin-flipping protocol, where neither side can 
cheat by a finite amount without getting caught, is described in
Ref.~\cite{Hardy99}, and is a strong coin-flipping protocol that is
constructed out of bit-commitment.

For the proof of the above statements, we assume the existence of a quantum
protocol $\mathcal{P}$ described by Eq.~(\ref{eq:Pc}) with $b=1$ and some
non-zero $a$. We shall describe a game, that uses $\mathcal{P}$ as a black
box, which achieves strong coin-flipping with bias that becomes arbitrarily
small in the limit of a game parameter $N\rightarrow\infty$. Though, for the
purposes of comparing against Kitaev's bound, it is sufficient to allow
honest players to output ``cheat'' at the end, we will construct the game
so that the honest player always outputs one of the outcomes zero or one. 

The game is the random walk on a 1-D line, starting from the origin,
using the coin provided by $\mathcal{P}$. The game ends with the first arrival
at one of the two sites $\pm N$, with the right end corresponding to the 
outcome zero, and the left end to one.

If one party detects cheating, they will continue the game using
a private fair coin, and output according to the outcome of the game.
Let $z\in\{-N,\dots,N\}$ be a variable that runs along the line. If cheating
occurs when the game is at $z$, then the honest players can simply output
using $P_0=(N+z)/2N$ and $P_1=(N-z)/2N$. In essence, the only deterrent to the
cheater is that he may be able to cheat more effectively in a future round.
Note that the honest party cannot
just flip a balanced coin after detecting cheating because in this case
the cheating party would only cheat when he is about to lose. 

Because the honest player keeps no state beyond the current location
along the line, $z$, there is an optimal cheating strategy where
the bias only depends on $z$. That is, we only need to consider functions
$\epsilon(z)$ when maximizing over cheating strategies.

We assume that the cheating player is trying to bias toward zero (i.e.,
the right side). We can then define the function $W_\epsilon(z)$ to be the 
probability of winning starting from node $z$, using biases $\epsilon(z)$. 
The function is similar to $P_W$ defined in the previous section, except that 
we are now using large biases, which cannot be ignored in calculating the
probability of winning.

The function has the constraints $W_\epsilon(N)=1$, $W_\epsilon(-N)=0$ and
\be
W_\epsilon(z) &=& P_0(\epsilon(z)) W(z+1) + P_1(\epsilon(z)) W(z-1)
\nonumber\\
&&+ P_C(\epsilon(z)) \frac{N+z}{2N}.
\label{eq:rec}
\ee

\noindent
Clearly, for optimal strategies, $W_\epsilon(z)\geq (N+z)/2N$ for all $z$
because the cheating party can always play honestly.

The value of $W_\epsilon(0)$ is simply the probability of winning
the entire game. The cheater will chose $\epsilon(z)$ in order to maximize
$W_\epsilon(0)$. We shall prove that
$\max\lp[W_\epsilon(0)-1/2\rp]\rightarrow 0$ as $N\rightarrow\infty$.

The analysis is made easier by using a modified black box protocol 
$\mathcal{P}'$ with achievable probabilities of the form:
\be
P_0 &=& \frac{1}{2} + \epsilon,\\
P_1 &=& \frac{1}{2} - (1+a) \epsilon,\\
P_C &=& a \epsilon,
\ee

\noindent
for $0\leq\epsilon\leq \epsilon_{max}$ where $\epsilon_{max}=1/(2+2a)$. 
For every $\epsilon$, protocol $\mathcal{P}'$ gives the cheater a slightly
higher probability of obtaining the desired outcome than with protocol 
$\mathcal{P}$. Therefore, any bounds on cheating obtained using $\mathcal{P}'$
as a black box will apply when using $\mathcal{P}$ instead.

Consider using $\mathcal{P}'$ and varying independently each of the nodes
of the complete game tree. The bias $\epsilon(x)$ of each node enters linearly
into $W_\epsilon(0)$, which can be maximized by letting the biases take
only the boundary values of $0$ or $\epsilon_{max}$. As discussed above,
the optimal biases will depend only on the corresponding value of $z$, and 
therefore we need only consider functions $\epsilon(z)$ which take values
in $\lp\{0,\epsilon_{max}\rp\}$.

The maximization is now easy to analyze. Define 
$\delta(z) = W_\epsilon(z)-(N+z)/2N$, which is the extra probability
of winning that the cheater is getting at position $z$. We shall prove 
below that
\be
\delta(z+1)\leq \frac{2+a}{2a N}
\ \ \ \Longrightarrow \ \ \ 
\delta(z)\leq \frac{2+a}{2a N}.
\label{eq:cond}
\ee

If $\epsilon(z)=\epsilon_{max}$ the above statement is true because 
Eq.~(\ref{eq:rec}) states that:
\be
\delta(z) &=& \lp(\frac{1}{2} + \epsilon_{max}\rp)
\lp(\delta(z+1) + \frac{1}{2N}\rp)\\\nonumber
&=& \frac{2+a}{2+2a}
\lp(\delta(z+1) - \frac{2+a}{2a N}\rp)+\frac{2+a}{2a N}.
\label{eq:emax}
\ee

On the other hand, if $\epsilon(z)=0$ then Eq.~(\ref{eq:rec}) states that:
\be
\delta(z+1)-\delta(z)=\delta(z)-\delta(z-1),
\ee

\noindent
that is, $\delta$ has a constant slope around $z$. 
If for all $z'<z$ we have $\epsilon(z')=0$, then the
slope must be constant through this entire region. Since $\delta(-N)=0$, the 
slope can only be negative (i.e. increasing toward the left)
if $\delta(z)<0$, which proves Eq.~(\ref{eq:cond})
for this case. Otherwise, let $z'<z$ be the largest integer such that
$\epsilon(z')=\epsilon_{max}$. Again the slope must be constant from 
$z'$, in which case, by Eq.~(\ref{eq:emax}) the slope can only be negative
if $\delta(z')<(2+a)/2a N$ which implies $\delta(z)<(2+a)/2a N$.

Having proven Eq.~(\ref{eq:cond}), and using the initial case
$\delta(N)=0$, we have shown
\be
\epsilon_{tot} \equiv W_\epsilon(0)-\frac{1}{2} 
= \delta(0) \leq \frac{2+a}{2a N},
\ee

\noindent
which can be made arbitrarily small by taking the limit $N\rightarrow \infty$.

The game described in this section shows that a strong coin-flipping protocol
with linear cheat detection can be serially composed to obtain a strong
coin-flipping protocol with arbitrarily small bias. The conclusion is that
quantum strong coin-flipping protocols with linear cheat detection cannot 
exist.

\section{Conclusions}

Using serial composition of coin-flipping we have established an
upper bound on the amount of cheat detection possible in quantum 
protocols for coin-flipping and bit-commitment. We have also presented evidence
that serially composing quadratic cheat sensitive protocols does not lead to
an improvement in the amount of cheat detection. We speculate that quadratic
or worse cheat sensitivity ($b\geq2$) cannot be composed in series to
obtain weak coin-flipping with arbitrarily small bias. We also speculate
that cheat detection better than quadratic ($b<2$) does not exist for
bit-commitment or strong coin-flipping, and probably not for weak
coin-flipping either.

Nevertheless, 
linear cheat detection in weak coin-flipping remains an open, 
though unlikely, possibility. In fact, by serially composing weak coin-flipping
with linear cheat detection, it is not hard to show that one can construct
a weak coin-flipping protocol with a bias of exactly zero. The apparent 
contradiction with the result of Lo and Chau \cite{Lo:1998pn} is resolved
by noting that they only considered protocols with a fixed number of rounds.
However, there are protocols where the number of rounds is variable and
possibly arbitrarily large (a good classical example is rock-paper-scissors),
while still having a zero probability of going on forever. For these protocols
the measurements cannot be delayed to the final round, and therefore the 
analysis of Lo and Chau does not apply. These protocols can always
be truncated to a finite number of rounds, though, at the cost of 
allowing an arbitrarily small bias.

Other questions that remain open include: What happens to serial composition
of quadratic cheat detection in the large bias regime? What can be said
for other functional forms of cheat detection versus bias, including cases
where one party may be a able to cheat by a small amount without getting
caught? And of course, the main question---whether or not weak coin-flipping 
with arbitrarily small bias can be achieved as a quantum protocol---remains 
unresolved.

\begin{acknowledgments}

This paper was written to provide an alternative to Canadian coins which have
an experimentally determined bias of $0.5$. The author would like to thank
John Preskill, Michael Ben-Or, Rob Spekkens and Ben Toner for their help.

This work was supported in part by
the National Science Foundation under grant number EIA-0086038 and
by the Department of Energy under grant number DE-FG03-92-ER40701.

\end{acknowledgments}

\appendix

\section{Proof of Lemma}

Let $\mathcal{T}$ be a binary tree. We will use the variable $x$ to denote
a binary node of $\mathcal{T}$ and $y$ to denote a leaf.

Associated with $\mathcal{T}$ there is a function $P_W(y)$ on the leaf nodes
which takes the values zero or one. The function can be extended to 
the rest of the nodes by defining $P_W(x)$ as the average of the 
function on its two descendants. In terms of coins, $P_W$ is simply the 
probability of winning when playing honestly starting from the given node. 
The function corresponds to the coin outcomes on the leaf nodes if the goal 
is to obtain $1$, and otherwise the coin outcome is $1-P_W(y)$.
To obtain a fair coin toss, we require that $P_W$ equal $1/2$ on the root 
node. Otherwise, the function and the tree are arbitrary.

We also arbitrarily label the two outgoing edges from each binary node as 
up and down, and define $\Delta(x)=P_W(x^\uparrow)-P_W(x^\downarrow)$. 
Finally, let $D(x)$ be the depth of node $x$, with the root node having
depth zero.

\textbf{Lemma:} Given a tree $\mathcal{T}$ and function $P_W$ as described
above, the following equality holds:
\be
\sum_x 2^{-D(x)} \Delta(x)^2 = 1.
\label{eq:lemma}
\ee

\noindent
Proof: $\Delta(x)$ is a linear combination of $P_W(x)$ on its
descendants, which in turn is a linear combination of $P_W(y)$ on the leaf
nodes. Therefore the left-hand side of the above equation is a quadratic
polynomial of $P_W(y)$. 

Fix a leaf node $y$. The function $P_W(y)$ only appears in $\Delta(x)$ if
$y$ is a descendant of $x$, in which case it has a coefficient of
$\pm 2^{D(x)+1-D(y)}$. The coefficient of $P_W(y)^2$ in this polynomial
is therefore given by
\beq
\sum_{i=0}^{D(y)-1} 2^{-i} 2^{2(i+1-D(y))} = 
2^{2(1-D(y))} \lp( 2^{D(y)} - 1\rp).
\eeq

\noindent
Fix a second leaf node $y'\neq y$. Let $x'$ be the unique node that has
$y$ as a descendant along one branch and $y'$ along the other.
The coefficient of $P_W(y) P_W(y')$ is
\be
&&\!\!\!\!\!\!
2\lp[\sum_{i=0}^{D(x')-1} 2^{-i+2(i+1)-D(y)-D(y')}\rp]
 - 2^{D(x')-D(y)-D(y')+3} \nonumber\\
&&\!\!\!\!\!\!
= -2^{3-D(y)-D(y')},
\ee

\noindent
where the only negative term is contributed by $\Delta(x')$. Note the
extra factor of $2$ accounting for the double occurrence of $P_W(y) P_W(y')$.

Combining these results, the left-hand side of Eq.~(\ref{eq:lemma}) is
\be
4\sum_y 2^{-D(y)} P_W(y)^2 - 4 \lp[ \sum_y 2^{-D(y)} P_W(y) \rp]^2.
\ee

\noindent
Note that the factor in brackets is just $P_W$ at the root node, which must
equal $1/2$. The left summand can be simplified by using $P_W(y)^2 =P_W(y)$,
in which case it can also be written in terms of $P_W$ at the root node.
We have shown that the left-hand side equals $4(1/2)-4(1/2)^2=1$ as
desired.$\Box$


\end{document}